\documentclass[prd,preprint,superscriptaddress,showpacs,amsmath,amssymb]{revtex4}
\usepackage{graphicx}
\usepackage{dcolumn}
\usepackage{bm}

\begin{document}

\title{Asymptotic quasinormal modes of scalar field in a gravity's rainbow}

\author{Cheng-Zhou Liu}
  \email{czlbj20@yahoo.com.cn}
  \affiliation{Department of Physics, Beijing Normal University, Beijing 100875, China}
  \affiliation{Department of Physics and Electronic Science, Binzhou College, Shandong 256600, China}

\author{Jian-Yang Zhu }
  \email{zhujy@bnu.edu.cn}
  \affiliation{Department of Physics, Beijing Normal University, Beijing 100875, China}

\begin{abstract}
In the context of a gravity's rainbow, the asymptotic quasinormal
modes of the scalar perturbation in the quantum modified
Schwarzschild black holes are investigated. By using the monodromy
method, we calculated and obtained the asymptotic quasinormal
frequencies, which are dominated not only by the mass parameter of
the spacetime, but also by the energy functions from the modified
dispersion relations. However, the real parts of the asymptotic
quasinormal modes is still $T_H\ln 3$, which is consistent with
Hod's conjecture. In addition, for the quantum corrected black hole,
the area spacing is calculated and the result is independent of the
energy functions, in spite of the area itself is energy dependence.
And that, by relating the area spectrum to loop quantum gravity, the
Barbero-Immirzi parameter is given and it remains the same as from
the usual black hole.

\end{abstract}

 \maketitle

\section{Introduction}

As is well known, the response of black hole background to the
initial configuration is dominated by the quasinormal modes(QNMs)-a
set of complex frequencies which only depend on the parameters of
the black hole rather than the initial perturbation (see \cite{1,2}
for reviews). The characterization of QNMs can be play a important
role in the observation of black hole in the university. Meanwhile,
the properties of QNMs can be used to investigate the black hole
stability. In addition, QNMs has get interpretation in Conformal
Field Theory \cite{3,4,5,6,7,8}. And that, the asymptotic QNMs can
be related to the quantum theory of black hole and quantum geometry
\cite{9,11}. That is, motivated by Hod's conjecture \cite{9}, it is
suggested that the value of asymptotic QNMs may carry important
information about the quantization of black hole area and help to
fix certain parameters in quantum geometry \cite{11}.

Based on the numerically results of QNMs \cite{12}, Hod suggested
that the real parts of highly damped QNMs in the Schwazchild black
hole can be expressed as $\omega _R=T_H\ln 3$ \cite{9}. And that, by
using Bohr's correspondence principle, the real part could be
identified as the characteristic transition frequency for the black
hole and the fundamental quanta of black hole area as $l_p^24\ln 3$
can be obtained \cite{9,11}. On the other hand, as a main candidate
theory of quantum gravity, loop quantum gravity(LQG) (see
\cite{27,28} for reviews) has a remarkable result-the discrete area
spectra. However, there is an unknown 'natural constant' called the
Barbero-Immirzi parameter and the area spectra have an ambiguity.
Using the area quanta of black hole, Dreyer presented the free
parameter and proposed that LQG should be based on $SO(3)$ rather
than $SU(2)$ \cite{11}. Thus, one hope that the asymptotic QNMs will
have a role in the quest for quantum geometry, especially for
quantum properties of black hole.

For analytically compute the asymptotic value of the QNMs and
confirm the Hod's conjecture, the monodromy method of investigating
quasinormal modes was proposed \cite{14}. In the method, the radial
coordinate is analytically continued into the complex plane and the
properties of the potential and the tortoise coordinate around the
origin, the horizons and the infinity are used \label{12}. The
obtained asymptotic QNMs present an analytical proof on the Hod's
conjecture for scalar and some gravitational perturbations in
Schwarzschild black hole. In the lectures \cite
{14,15,16,17,18,19,20,nbl,21,22}, the monodromy technique has been
extend to many cases such as different perturbations like Dirac
field and electromagnetic field in more general static black holes
and the asymptotic QNMs have been obtained. However, the researches
imply that whether the argument $\omega _R=T_H\ln 3$ applies to
universal cases is still a open problem and much less attention was
paid to the case of analytically calculating the asymptotic QNMs in
quantum corrected black hole \cite{nbl}.

It is believed that, with quantum gravity effects, a discrete
picture of spacetime will emerge and a new physics will be
necessary. In \cite{n-1,30}, the Planck scale corrected spacetime
named as gravity's rainbow has been presented. The main feature of
the gravity's rainbow is that the geometry of spacetime depend on
the energy of a particle moving in it. That is to say, for the
spacetime with Planck scale correction effects, there are different
geometries for probe particles with different energies. The modified
geometries of spacetime can be described by one parameter family of
metric as a function of particle's energy observed by an inertial
observer. Moreover, the modified Schwarzschild solution in the
gravity's rainbow has been given \cite{30} and it's some
thermodynamics quantities and asymptotic flatness have been
investigated \cite{31,31-1}.

Here, we investigated the asymptotic QNMs of massless scalar field
in the modified Schwarzschild spacetime from the gravity's rainbow.
Our main aim is to verify the influence of spacetime's quantum
effects on the asymptotic QNMs and to test the Hod's conjecture and
it's implies in the quantum corrected spacetime. By using the
monodromy method, we analytically calculate the asymptotic
quasinormal frequencies in the quantum corrected spacetime. The
results show that, when the Plank scale modification of spacetime is
taken into account, the asymptotic QNMs depend on not only the mass
parameter of the black hole but also the energy functions, by which
the quantum effects of spacetime are reflected. However, the real
part of the asymptotic QNMs can still be expressed as $T_H\ln 3$ and
hence the Hod's conjecture is valid for the gravity's rainbow. In
addition, in the quantum corrected spacetime, some quantum implies
\cite{9,11} of the Hod's conjecture are verified. The area spacing
of the quantum corrected black hole is calculated and the free
parameter of LQG is presented. The obtained results are independent
of the energy functions and remain the same as from the usual
Schwarzschild black hole \cite{9,11}.

The paper is organized as follows. In Sec. \ref{Sec.2}, the modified
Schwarzschild solution from the gravity's rainbow is introduced
briefly. Then in Sec. \ref{Sec.3}, in the monodromy method
framework, the asymptotic QNMs of massless scalar field in the
quantum corrected spacetime are calculated and obtained
analytically. Section \ref{Sec.4}, for the quantum corrected black
hole, Hod's conjecture and it's implies in quantum theory of black
hole and LQG are verified. The last part \ref{Sec.5} is the summary
and conclusions.

\section{Gravity's rainbow and Modified Schwarzschild black holes}

\label{Sec.2}Let us firstly briefly introduce the modified
Schwarzschild solution from the gravity's rainbow. When keeping
Planck energy as an invariant scale, namely a universal constant for
all inertial observers, to preserve the relativity of inertial
frames, the double special relativity (DSR) has been proposed
\cite{23-1,23-2,23-3,23,24}. The staring point and the main result
of DSR is the modified dispersion relation (MDR) as
\begin{equation}
E^2f_1^2\left( E{;\lambda }\right) -p^2f_2^2\left( E{;\lambda
}\right) =m_0^2,  \label{eq1}
\end{equation}
where $f_1$ and $f_2$ are two energy functions from which rotational
symmetry can be preserved, $\lambda $ is a parameter of order the
Planck scale. The equation Eq.(\ref{eq1}) shows that, MDR is energy
dependent. It is to say, particles with different energy $E$ have
different energy-momentum relations. However, in the low energy
realm i.e. $E/E_p{<<1}$, $f_1$ and $f_2$ approach to unit and MDR
can return to the usual energy-momentum relation, where $E_p\equiv
1/\sqrt{8\pi G}$ is the Planck energy. This is consistent with
Bohr's correspondence principle.

In the context of DSR, the deformed spacetime geometry has been
investigated and some proposals have been presented
\cite{n-1,30,3-1,3-2,3-3,3-4}. By the \cite{n-1,30}, it is put
forwarded that the flat spacetime with Planck scale corrections has
the invariant as
\begin{equation}
ds^2=-\frac{dt^2}{f_1^2}+\frac{dr^2}{f_2^2}+\frac{r^2}{f_2^2}d\Omega
^2. \label{eq2}
\end{equation}
This equation indicates that, the DSR spacetime depend on the energy
of particle moving in it. That is to say, particles with different
energy will probe different DSR spacetimes. Thus, the DSR spacetime
is endowed with an energy dependent quadratic invariant, that is, an
energy dependent metric, namely rainbow metric. Just like the DSR,
when particle's energy is a little quantity comparing with the
Planck energy, the rainbow metric will turn to the usual flat
spacetimes. And that, it has been pointed that the DSR spacetimes
has equality with the usual flat spacetimes \cite{31-1}.

By extending the Eq.(\ref{eq2}) to incorporate curvature, in
\cite{30}, a gravity's rainbow has been presented and a few quantum
corrected spacetimes from the gravity's rainbow have been given.
Thereamong, the modified Schwarzschild solution can be expressed in
terms of energy independent coordinates and the energy independent
mass parameter as
\begin{equation}
dS^2=-\frac{\left( {1-}\frac{2GM}r\right) }{f_1^2}dt^2+\frac 1{f_2^2\left( {%
1-}\frac{2GM}r\right) }dr^2+\frac{r^2}{f_2^2}d\Omega ^2.
\label{eq3}
\end{equation}
Obviously, the metric of the modified Schwarzschild black hole is
also energy dependent. That is, if a given observer probes the
spacetime using the quanta with different energies, he will conclude
that spacetime geometries have different effective description.
Here, the particle's energy $E$ denotes it's total energy measured
at infinity from the black hole. In addition, the energy dependence
of the gravity's rainbow probably may has relation to the property
of background independent of LQG \cite{27,28}. In fact, as a
semi-classical limit, the MDR has been suggested by LQG \cite
{DL-3,DL-4,DL-1,DL-2}. Thus, as a gravity's rainbow, the present
spacetime is endowed with Plank-scale modifications, i.e., quantum
effects.

\section{Asymptotic quasinormal modes}

\label{Sec.3}

Now, in the monodromy method framework \cite{14}, we analytically
calculate the asymptotic QNMs of massless scalar field in the
modified Schwarzschild
from the gravity's rainbow. Firstly, as in the usual Schwarzschild black hole%
\cite{32}, by using the Klein-Gordon equation, we give the
perturbation equation for the scalar field $\phi $ in the quantum
corrected spacetime. Then, substituting Eq.(\ref{eq3}) into the
Klein-Gordon equation
\begin{equation}
\frac 1{\sqrt{-g}}\frac \partial {\partial x_\mu }\left(
\sqrt{-g}g^{\mu \nu }\frac \partial {\partial x_\nu }\right) \phi
=0,  \label{eq10}
\end{equation}
and using the ansatz of separation of variables $\phi =\frac
1re^{i\omega t}\psi (r)Y(\theta ,\varphi )$, the radial perturbation
equation can be given as
\begin{equation}
\left( r^2-2GMr\right) \frac{\partial ^2\psi (r)}{\partial r^2}+2GM\frac{%
\partial \psi (r)}{\partial r}+\left[ r^2\frac{f_2^2}{f_1^2}\left( 1-\frac{%
2GM}r\right) ^{-1}\omega ^2-\frac{2GM}{r^3}-\frac{l\left( l+1\right) }{r^2}%
\right] \psi (r)=0,  \label{eq10-2}
\end{equation}
where $l$ is the orbital angular momentum.

For convenience, we define the tortoise coordinate of the quantum
modified black hole as
\begin{equation}
x=\int \frac{dr}{h\left( r\right) }=\frac{f_1}{f_2}\left[ r+2GM\ln
\left( \frac r{2GM}-1\right) \right] ,  \label{eq11}
\end{equation}
with
\begin{equation}
h\left( r\right) =\frac{f_2}{f_1}\left( 1-\frac{2GM}r\right) .
\label{q11-1}
\end{equation}
Obviously, at the infinity, we have $x\rightarrow +\infty $. And
that, in the vicinity of $r=0$, we have $h\left( r\right) \sim
-\frac{f_2}{f_1}\frac{2GM}r $ and the tortoise coordinate takes the
form
\begin{equation}
x\sim -\int \frac{f_1}{f_2}\frac
r{2GM}dr=-\frac{f_1}{f_2}\frac{r^2}{4GM}. \label{eq11-3}
\end{equation}
In addition, from the metric Eq.(\ref{eq3}), it is seen that the
event horizon of the modified black hole is at
\begin{equation}
r_{+}=2GM.  \label{eq5}
\end{equation}
Then, near the event horizon, we have $h\left( r\right) \sim \left(
r-r_{+}\right) h^{^{\prime }}\left( r_{+}\right) =\frac{f_2}{f_1}\frac{r-2GM%
}{2GM}$ and $x\sim \frac{f_1}{f_2}2GM\ln \left( r-2GM\right) $. This
shows that, at the horizon, we have $x\rightarrow -\infty $.

Thus, substituting Eq.(\ref{eq11}) into Eq.(\ref{eq10-2}), the
radial wave equation is translated into
\begin{equation}
\frac{d^2\psi (r)}{dx^2}+\left[ \omega ^2-V\left( r\right) \right]
\psi (r)=0,  \label{eq12}
\end{equation}
where $V\left( r\right) $ is the effect potential with
\begin{equation}
V\left( r\right) =\frac{f_2^2}{f_1^2}\left( 1-\frac{2GM}r\right)
\left( \frac{l\left( l+1\right) }{r^2}+\frac{2GM}{r^3}\right) .
\label{eq13}
\end{equation}
We see that, in the quantum corrected black hole, the effective
potential depend not only the radial coordinate and the angular
quantum number but also the energy functions. This shows that the
effective potential has some
Plank-scale modifications. But, as in the usual black hole \cite{13}, while $%
x\rightarrow \pm \infty $, we have $V\left( r\right) \rightarrow 0$,
then the Eq.(\ref{eq12}) can be translated into the standard wave
equation in terms of the tortoise coordinate Eq.(\ref{eq11}). That
is, the solutions of Eq.(\ref{eq12}) behave as $\psi (x)\sim e^{\pm
i\omega x}$ at the horizon and infinity.

As the same in the usual black hole, the QNMs in the modified black
hole can be supposed to be the solutions of perturbation equation
Eq.(\ref{eq12}) with the boundary conditions of purely ingoing waves
at event horizon and purely outgoing waves at infinity, namely
\begin{equation}
\psi (x)\sim e^{+i\omega x},x\rightarrow -\infty ,  \label{eq14}
\end{equation}
\begin{equation}
\psi (x)\sim e^{-i\omega x},x\rightarrow +\infty .  \label{eq15}
\end{equation}
Then, to analytically obtain the asymptotic QNMs for scalar field in
the modified Schwarzschild black hole, we calculate the differential
equation Eq.(\ref{eq12}) by using the monodromy method. In the
technology, it is
essential to extend analytically Eq.(\ref{eq12}) from it's physical region $%
r_{+}<r<\infty $ to the whole complex $r$-plane. By the extension,
it is find that the solutions $\psi (r)$ behave multivaluedness
around the singular points $r=0$ and $r=r_{+}$ and that the
asymptotic QNMs will be obtained by computing the monodromy around a
chosen contour in the complex plane. As in
\cite{14}, we put a branch cut from $r=0$ to $r=r_{+}$ and the monodromy of $%
\psi (r)$ can be defined by the discontinuity across the cut.
Moreover, from Eq.(\ref{eq11}), the Stokes line $%
\mathop{\rm Re}%
\left( x\right) =0$ in the modified black hole can be obtained and
the contour $L$ can be chosen as in figure 1.

We see that, inside the contour, the only singularity of $\psi (r)$
is the event horizon. According to the boundary condition
Eq.(\ref{eq14}) and the
tortoise coordinate Eq.(\ref{eq11}), we can obtain that the monodromy of $%
\psi (r)$ around the contour $L$ is the same as the monodromy of
$e^{i\omega x}$ and must be\cite{14}
\begin{equation}
e^{\frac{\pi \omega }k},  \label{eq16}
\end{equation}
where $\kappa $ is the surface gravity on the event horizon. For the
modified black hole shown as Eq.(\ref{eq3}), $\kappa $ can be
obtained by
\begin{equation}
\kappa =-\frac 12\lim_{r\rightarrow
r_{+}}\sqrt{\frac{-g^{rr}}{g^{tt}}}\frac 1{g^{tt}}\frac{\partial
g^{tt}}{\partial r}=\frac{f_2}{f_1}\frac 1{4GM}. \label{eq7}
\end{equation}
Then, the Hawking temperature can be given as
\begin{equation}
T_H=\frac \kappa {2\pi }=\frac{f_2}{f_1}\frac 1{8\pi GM}.
\label{eq9}
\end{equation}
We see that the surface gravity and the Hawking temperature of the
modified Schwarzschild black hole are all depend on the energy of
probe particle. That is, using the quanta with different energy, an
observer at infinity will probe different effective temperature and
surface gravity for the quantum corrected black hole. It is easy to
verify that, in the context of the gravity's rainbow, the energy
dependence of surface gravity and the temperature ought to arises
from MDR and should be the exhibition of quantum effects of the
spacetime. And that, this should have some modification effects on
the black hole physics, for example the QNMs.

Next, we compute the local monodromy of $\psi (x)$ around the same
contour. For $r\sim 0$, from Eqs.(\ref{eq11-3}) and (\ref{eq13}), we
have
\begin{equation}
V\left( r\right) \sim -\frac 1{4x^2}.  \label{eq18}
\end{equation}
It is found that, in the vicinity of $r=0$, the potential has the
simply form as in the usual Schwarzschild black hole \cite{14}.
Here, we could consider that the energy dependence of $V\left(
r\right) $ is involved in the tortoise coordinate.

Similarly to \cite{14}, for using Bessel's equation, we first
rewrite the potential to the form
\begin{equation}
V\left( r\right) \sim -\frac{1-j^2}{4x^2},  \label{eq19}
\end{equation}
where $j$ is the spin of the perturbation field. For our
interesting, the
results for the scalar perturbation could be obtain from the case $j=0$. Thus, near $%
r=0 $, we obtain the perturbation equation
\begin{equation}
\left( \frac{\partial ^2}{\partial x^2}+\omega
^2+\frac{1-j^2}{4x^2}\right) \psi (x)=0,  \label{eq20}
\end{equation}
and it's solutions can be expressed as \cite{14}
\begin{equation}
\psi (x)\sim B_{+}\sqrt{2\pi \omega x}J_{\frac j2}\left( \omega
x\right) +B_{-}\sqrt{2\pi \omega x}J_{-\frac j2}\left( \omega
x\right) ,  \label{eq21}
\end{equation}
where $B_{\pm }$ are integral constants and $J_\nu $ represents the
Bessel function of order $\nu .$

For the asymptotic QNMs, $\omega $ is approximately a purely
imaginary and
hence the line Im$\left( \omega x\right) =0$ is almost the same as the line $%
\mathop{\rm Re}%
\left( x\right) =0$. Then, on the contour $L$, $x\rightarrow
+\infty $ can actually rotated to $\omega x\rightarrow +\infty $ and
the boundary condition Eq.(\ref{eq15}) can be expressed as
\begin{equation}
\psi \sim e^{-i\omega x},\omega x\rightarrow +\infty .  \label{eq22}
\end{equation}
And that, along the contour $L$, it is convenient to obtain the
asymptotic forms of $\psi (x)$ away from the origin $r=0$ to $\omega
x\rightarrow +\infty $. For $z\gg 1,$ considering the asymptotic
behavior
\begin{equation}
J_\nu \left( z\right) =\sqrt{\frac 2{\pi z}}\cos \left( z-\frac{\nu \pi }%
2-\frac \pi 4\right) ,  \label{eq23}
\end{equation}
we obtain
\begin{eqnarray}
\psi (x) &\sim &2B_{+}\cos \left( \omega x-\alpha _{+}\right)
+2B_{-}\cos
\left( \omega x-\alpha _{-}\right)  \nonumber \\
&=&\left( B_{+}e^{-i\alpha _{+}}+B_{-}e^{-i\alpha _{-}}\right)
e^{i\omega x}+\left( B_{+}e^{i\alpha _{+}}+B_{-}e^{i\alpha
_{-}}\right) e^{-i\omega x}, \label{eq24}
\end{eqnarray}
where the phase shifts $\alpha _{\pm }=\frac \pi 4(1\pm j)$. Thus,
from the boundary condition Eq.(\ref{eq22}), it is seen that, as
$\omega x\rightarrow +\infty $, the $B_{\pm }$ must satisfy
\begin{equation}
B_{+}e^{-i\alpha _{+}}+B_{-}e^{-i\alpha _{-}}=0,  \label{eq25}
\end{equation}
and the solution $\psi (x)$ take the form

\begin{equation}
\psi (x)\sim \left( B_{+}e^{i\alpha _{+}}+B_{-}e^{i\alpha
_{-}}\right) e^{-i\omega x}.
\end{equation}
Now, along the contour $L$, let us from point $A$ approach to point
$B$. In the process, we must turn an angle $3\pi /2$ around the
origin, i.e., $3\pi $ around $x=0$. As well known, for $z\sim 0$,
the Bessel functions satisfy
\begin{equation}
J_\nu \left( z\right) =z^\nu w\left( z\right) ,  \label{eq27}
\end{equation}
where $w\left( z\right) $ is an even holomorphic function. Then,
after the rotation, we have
\begin{equation}
\sqrt{2\pi e^{i3\pi }\omega x}J_{\pm \frac j2}\left( e^{i3\pi
}\omega x\right) \sim 2e^{6i\alpha _{\pm }}\cos \left( -\omega
x-\alpha _{\pm }\right) ,  \label{eq28}
\end{equation}
and the solutions for $\omega x\rightarrow -\infty $ can be obtained
as
\begin{eqnarray}
\psi (x) &\sim &2B_{+}e^{6i\alpha _{+}}\cos \left( -\omega x-\alpha
_{+}\right) +2B_{-}e^{6i\alpha _{-}}\cos \left( -\omega x-\alpha
_{-}\right)
\nonumber \\
&=&\left( B_{+}e^{7i\alpha _{+}}+B_{-}e^{7i\alpha _{-}}\right)
e^{i\omega x}+\left( B_{+}e^{5i\alpha _{+}}+B_{-}e^{5i\alpha
_{-}}\right) e^{-i\omega x}.  \label{eq29}
\end{eqnarray}
Next, we continue along the contour $L$ in the right half-plane. In
the large semicircle of $L$, the term $\omega ^2$ dominates the
potential $V$ and hence the solutions $\psi (x)$ of the perturbation
equation Eq.(\ref {eq20}) can be approximated as plane waves. By
this way, when we complete the contour, the coefficient of
$e^{-i\omega x}$ in the $\psi (x)$ can be remain unchanged and the
coefficient of $e^{i\omega x}$ makes only an exponentially small
contribution to the solutions. Thus, when we return to the point $A$
from $B$, the coefficient of $e^{-i\omega x}$ gets multiplied by
\begin{equation}
\frac{B_{+}e^{5i\alpha _{+}}+B_{-}e^{5i\alpha _{-}}}{B_{+}e^{i\alpha
_{+}}+B_{-}e^{i\alpha _{-}}}.  \label{eq30}
\end{equation}
In addition, similarity to Eq.(\ref{eq16}), the monodromy of
$e^{-i\omega x}$
around the contour is $e^{-\frac{\pi \omega }k}$. Thus, for the solutions $%
\psi (x)$, to have the required monodromy of Eq.(\ref{eq16}), we get
\begin{equation}
\frac{B_{+}e^{5i\alpha _{+}}+B_{-}e^{5i\alpha _{-}}}{B_{+}e^{i\alpha
_{+}}+B_{-}e^{i\alpha _{-}}}e^{-\frac{\pi \omega }k}=e^{\frac{\pi
\omega }k}. \label{eq31}
\end{equation}
Considering Eq.(\ref{eq25}), the above equation can be translated
into
\begin{equation}
e^{\frac{2\pi \omega }k}=\frac{e^{6i\alpha _{+}}-e^{6i\alpha _{-}}}{%
e^{2i\alpha _{+}}-e^{2i\alpha _{-}}}=-\left( 1+2\cos \pi j\right) .
\label{eq32}
\end{equation}
Therefore, letting $j=0$ and considering Eq.(\ref{eq7}), for scalar
perturbation in the modified Schwarzschild black hole from the
gravity's rainbow, the asymptotic QNMs are obtained as
\begin{equation}
\omega =\frac k{2\pi }\ln 3+ik\left( n+\frac 12\right) =\frac{f_2}{f_1}\frac{%
\ln 3}{8\pi GM}+i\frac{f_2}{f_1}\frac 1{4GM}\left( n+\frac 12\right)
, \label{eq33}
\end{equation}
where $n\rightarrow \infty $.

Comparing the asymptotic QNMs with the results from the usual Schwarzschild black hole%
\cite{14}, it is find that the present asymptotic quasinormal
frequencies depend not only on the mass parameters of the black
hole, but also on the energy functions $f_1$ and $f_2$. This show
that, the result obtained here has get quantum corrections exhibited
as the dependence on the energy of probe particle. However, in the
quantum corrected black hole, the real part of asymptotic
quasinormal frequencies is still $T_H\ln 3$ and hence consistent
with Hod's conjecture. And that, for the case of low energy, the
asymptotic quasinormal
frequencies can return to the value in the usual Schwarzschild black hole%
\cite{14}.

\section{Area spectrum}

\label{Sec.4}In the section, as for the usual black hole
\cite{9,11}, we verify the area spectrum of the quantum corrected
black hole and it's implies in LQG. Based on Bohr's correspondence,
Hod argue that \cite{9}, the process of highly damped QNMs of black
hole should be identical to the quantum transition of the
corresponding system and the real part of the asymptotic QNMs should
be equal to the quantum transition frequency. Then, in the modified
black hole, with the radiated wave shown as Eq.(\ref{eq33}), the
energy of the corresponding radiation quanta is obtained as

\begin{equation}
E=\hbar
\mathop{\rm Re}%
\left( \omega \right) =\frac{f_2}{f_1}\frac \hbar {8\pi GM}\ln 3.
\label{eq34}
\end{equation}

In other hand, from Eq.(\ref{eq3}), we can see the modified
Schwarzschild solution from the gravity's rainbow is asymptotically
DSR. And that, the asymptotically DSR spacetimes has equality with
the asymptotically flat spacetimes \cite{31-1}. Then, using the
Komar integrals, we define the total Arnowitt-Deser-Misner (ADM)
mass for the quantum corrected spacetime as
\begin{equation}
M_{ADM}=-\frac 1{8\pi G}\int_s{{\varepsilon _{abcd}\nabla ^c\xi
^d}}=\frac M{f_1f_2}.  \label{eq4}
\end{equation}
We find that, for the modified Schwarzschild black holes, the ADM
mass is not equal to the mass parameter $M$. This shows that the
quantum corrected spacetime have topological defects. Moreover, the
total energy of the spacetime depend on the energy of the probe
particle. Obviously, this is different from the usual Schwarzschild
black hole, in which $M_{ADM}=M$. And that, the energy property for
the modified black hole should arise from the quantum effects of the
spacetime, which is the gravity's rainbow with the energy
dependence.

Then, in the quantum modified black hole, after the radiation quanta
shown in Eq.(\ref{eq34}), the change of the mass parameter of the
black hole is obtained by

\begin{equation}
\triangle M=f_1f_2E=f_2^2\frac \hbar {8\pi GM}\ln 3.  \label{eq55}
\end{equation}

In addition, from Eqs.(\ref{eq3}) and (\ref{eq5}), the event horizon
area of the quantum corrected black hole can be obtained as
\begin{equation}
A=\int \left( g_{\theta \theta }g_{\varphi \varphi }\right)
_{r_{+}}d\theta d\varphi =\frac{16\pi G^2M^2}{f_2^2}.  \label{eq6}
\end{equation}
It should been pointed that, despite the event horizon
Eq.(\ref{eq5}) is universal for all observers and at the same place
as the usual Schwarzschild black hole, the horizon area of the
gravity's rainbow is energy dependence and different from the value
of the usual black hole.

Thus, for the quantum modified black hole, the area change
corresponding to the radiation quanta from the asymptotic QNMs is
obtained as

\begin{equation}
\triangle A=\frac 1{f_2^2}32\pi G^2M\triangle M=4\ln 3l_p^2,
\label{eq36}
\end{equation}
with the Planck length $l_p=\sqrt{G\hbar }$. We find that, for the
quantum modified black hole, the area spacing is the same as the
usual black hole \cite{9,11} and is all the same for all the
observers, despite the horizon area itself is different from the
value of the usual black hole and for different observers. This
should be investigated future.

As mentioned in In Sec.\ref{Sec.2}, the gravity's rainbow may has
some consistency with LQG. Next, as in \cite{11}, we relate the area
spectrum of the gravity's rainbow to quantum geometry. In LQG, the
area of a black hole is quantized and the area element is given
as\cite{29,27,28}
\begin{equation}
A(j_{\min })=8\pi l_p^2\sqrt{j_{\min }(j_{\min }+1)},  \label{eq37}
\end{equation}
where $\gamma $ is the so-called Barbero-Immirzi parameter and
$j_{\min }$ the smallest spin of the spin network intersecting the
event horizon. Then, under the argument \cite{11}
\begin{equation}
A(j_{\min })=\triangle A,  \label{eq37-1}
\end{equation}
the $\gamma $ can be expressed as
\begin{equation}
\gamma =\frac{\ln 3}{2\pi \sqrt{j_{\min }(j_{\min }+1)}}.
\label{eq38}
\end{equation}
In addition, black hole entropy from loop quantum gravity
is\cite{29-1,27,28}
\begin{equation}
S=N\ln (2j_{\min }+1),  \label{eq39}
\end{equation}
where $N=\frac A{A(j_{\min })}$ is the number of area element on the
event horizon. Then, considering Eq. (\ref{eq37-1}), we have
\begin{equation}
S=\frac A{4l_p^2}\frac{\ln (2j_{\min }+1)}{\ln 3}.  \label{eq40}
\end{equation}
Thus, letting the black hole entropy equal to the Bekenstein-Hawking
entropy as
\begin{equation}
S_{bh}=\frac A{4l_p^2},  \label{eq40-1}
\end{equation}
the smallest spin value is presented as
\begin{equation}
j_{\min }=1.  \label{eq41}
\end{equation}
The result is the same as obtained from the usual black hole and
suggested that LQG should be based on $SO(3)$ \cite{11}. And that,
substituting the spin value into Eq.(\ref{eq38}), the
Barbero-Immirzi parameter can be fixed as
\begin{equation}
\gamma =\frac{\ln 3}{2\pi \sqrt{2}}.  \label{eq}
\end{equation}

It is seen that, as the smallest spin value, the Barbero-Immirzi
parameter has not the dependence on the probe particle and as the
same as the results from the usual black hole \cite{11}. Obviously,
as a elementary parameter of LQG, it's independence on observer and
spacetime background is satisfactory.
\section{Summary and Conclusion}

\label{Sec.5}In the present work, the asymptotic QNMs in the quantum
modified Schwarzschild black holes from the gravity's rainbow are
investigated. By using the monodromy method, the asymptotic QNMs for
massless scalar field in the quantum corrected spacetime are
analytically calculated and obtained. The results show that the
highly damped quasinormal frequencies depend not only on the mass
parameter of the black hole, but also on the energy of probe
particle. Comparing with the results from the usual Schwarzschild
black hole, in which the asymptotic QNMs only depend on the mass
parameter \cite{14}, the energy dependence of the presented
asymptotic QNMs should be a quantum effect and come from the quantum
corrections of the background spacetimes. This is a novel property
of the asymptotic QNMs in the gravity's rainbow and should be proved
by other cases, such as different field in different gravity's
rainbow. It is also find that, in the context of gravity's rainbow,
the quantum effect of the asymptotic quasinormal frequencies is
involved in the Hawking temperature and the real part of asymptotic
QNMs is still $T_H\ln 3$. Thus, in support Hod's conjecture\cite{9},
the present reach provide further evidence from a quantum corrected
spacetime. In addition, by relating Hod's conjecture to quantum
theory of black hole \cite{9,11}, the area quantization of the
quantum modified black hole is investigated and the area spacing is
obtained. The result is energy independent and is the same as coming
from the usual Schwarzschild black hole \cite{9,11}, despite the
area itself is energy dependent and different from the value from
the usual black hole. And that, in the present quantum corrected
spacetime, by relating the area quantization to LQG \cite {11}, the
smallest spin value and the free parameter in LQG are given. Also,
the obtained results are energy independent and are the same as in
the usual black holes \cite{11}. The feature of the obtained area
spectrum and the elementary parameter being independent of observer
and background metric is in good agreement with expectation and
should be verified in future.

\acknowledgments The work was supported by the National Natural
Science of China (No.10375008) and the National Basic Research
Program of China (2003CB716302).

\newpage
\begin{figure}[tbp]
\includegraphics[clip,width=0.65\textwidth]{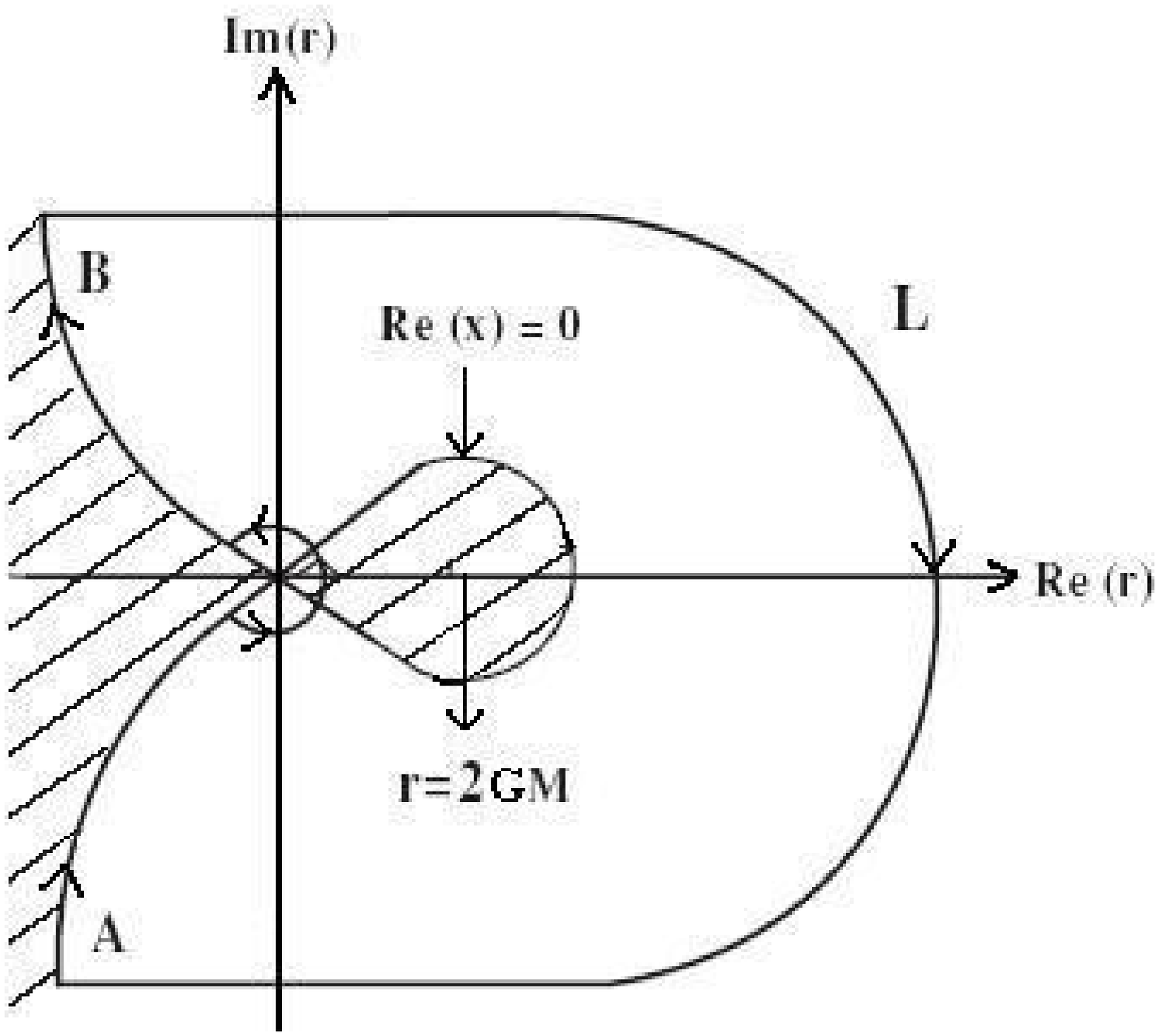}
\caption{The Stokes line $%
\mathop{\rm Re}%
\left( x\right) =0$ and the contour $L$ for the modified
Schwarzschild black hole from the gravity's rainbow. The
area of $%
\mathop{\rm Re}%
\left( x\right) <0$ is denoted by the regions with the hachures. }
\label{Fig.1}
\end{figure}


\begin{thebibliography}{99}
\bibitem{1}  Nollert H-P 1999 Quasinornmal Modes: The Characteristic ''Sound''
of Black Holes and Neutron Stars, Class. Quant. Grav. \textbf{16} R159

\bibitem{2}  Kokkotas K D and Schmidt B G 1999 Quasinornmal Modes of Stars and
Black Holes, Living Rev. Rel. \textbf{2} 2 arXiv:gr-qc/9909058

\bibitem{3}  Horowitz G T and Hubeny V E 2000 Phys. Rev. D \textbf{62} 024027

\bibitem{4}  Starinets A O 2002 Phys. Rev. D \textbf{66} 124013 arXiv:hep-th/0207133

\bibitem{5}  Cardoso V Konoplya R and Lemos J P S 2003 Phys. Rev. D \textbf{68}
044024 arXiv:gr-qc/0305037

\bibitem{6}  Birmingham D Sachs I and Solodukhin S N 2002 Phys. Rev. Lett.
\textbf{88} 151301 arXiv:hep-th/0112055

\bibitem{7}  Konoplya R A 2002 Phys. Rev. D \textbf{66} 044009 arXiv:hep-th/02050142

\bibitem{8}  Cardoso V and Lemos J P S 2001 Phys.Rev.D \textbf{63} 124015 arXiv:gr-qc/0101052

\bibitem{9}  Hod S 1998 Phys. Rev. Lett. \textbf{81} 4293 arXiv:gr-qc/9812002

\bibitem{11}  Dreyer O 2003 Phys.Rev.Lett. \textbf{90} 081301 arXiv:gr-qc/0211076

\bibitem{12}  Nollert H P 1993 Phys.Rev.D \textbf{747} 5253

\bibitem{27}  Ashtekar A and Lewandomski J 2004 Class Quantum Grav \textbf{21} R53-R152 arXiv:qr-qc/04040182

\bibitem{28}  Thiemann T 2001 arXiv:gr-qc/0110034

\bibitem{13}  Motl L 2002 Adv. Theor. Math. Phys. \textbf{6} 1135

\bibitem{14}  Motl L and Neitzke A 2003 Adv. Theor. Math. Phys. \textbf{7} 307

\bibitem{15}  Birmingham D 2003 Phys.Lett.B \textbf{69} 199

\bibitem{16}  Natario J and Schiappa R 2004 Adv. Theor. Math. Phys \textbf{8} 1001

\bibitem{17}  Ghosh A Shankaranarayanan S and Das S 2006 Class. Quant. Grav. \textbf{23} 1751

\bibitem{18}  Cardoso V Lemos J P S and Yoshida S 2004 Phys. Rev. D \textbf{69} 044004

\bibitem{19}  Lopez-Ortega A 2006 Gen. Rel. Grav \textbf{38} 1747

\bibitem{20}  Musiri S and Siopsis G 2006 arXiv:hep-th/0610170

\bibitem{nbl}  Giri P R 2006 arXiv:hep-th/0604188

\bibitem{21}  Cardoso V Natario J and Schiappa R 2004 J. Math. Phys. \textbf{45}
4698 arXiv:hep-th/0403132.

\bibitem{22}  Chen B S and Jing J L 2005 Class. Quant. Grav \textbf{22} 2159

\bibitem{n-1}  Kimberly D Magueijo J and Medeiros J 2004 Phys.Rev.D\textbf{70}
 084007 arXiv:gr-qc/0303067

\bibitem{30}  Magueijo J and Smolin L 2004 Class. Quant. Grav. \textbf{21} 1725 arXiv:gr-qc/0305055

\bibitem{31}  Ling Y Li X and Hu B 2005 arXiv:gr-qc/0512084

\bibitem{31-1}  Hackett J 2006 Class. Quant. Grav \textbf{23}3833

\bibitem{23-1}  Amelino-Camelia G 2002 Int. J. Mod. Phys. D \textbf{11} 35 arXiv:gr-qc/0012051

\bibitem{23-2}  Amelino-Camelia G 2001 Phys. Lett. B \textbf{510} 255 arXiv:hep-ph/0012238

\bibitem{23-3}  Kowalski-Glikman J 2001 Phys. Lett. A \textbf{285} 391

\bibitem{23}  Magueijo J and Smolin L 2002 Phys. Rev. Lett.\textbf{88} 190403 arXiv:hep-th/0112090

\bibitem{24}  Magueijo J and Smolin L 2003 Phys.Rev.D \textbf{67} 044017 arXiv:gr-qc/0207085

\bibitem{32}  Regge T and Wheeler J A 1957 Phys. Rev. \textbf{108} 1063

\bibitem{3-1}  Galan P and Mena Marugan G A 2004 Phys. Rev. D \textbf{70} 124003

\bibitem{3-2}  Galan P and Mena Marugan G A 2005 Phys. Rev.D \textbf{72} 044019

\bibitem{3-3}  Mignemi S 2003 Phys. Rev.D \textbf{68} 065029

\bibitem{3-4}  Hinterleitner F 2005 Phys. Rev. D \textbf{71} 025016

\bibitem{DL-3}  Gambini R and Pullin L 1999 Phys. Rev. D \textbf{59} 124021

\bibitem{DL-4}  Alfaro J Morales-Tecotl H A and Urrutia L F 2002 Phys. Rev. D
\textbf{65} 103509

\bibitem{DL-1}  Sahlmann H and Thiemann T 2002 arXiv:gr-qc/0207031

\bibitem{DL-2}  Smolin L 2005 arXiv:hep-th/051091

\bibitem{29}  Rovelli C and Smolin L 1995 Nucl. Phys. B \textbf{442} 593

\bibitem{29-1}  Rovelli C 1996 Phys. Rev. Lett. \textbf{77} 3288 arXiv:gr-gc/9603063
\end{thebibliography}
\end{document}